\batchmode
\documentstyle[12pt]{article}
\def\zid{1\kern-0.36em\llap~1}

\newcommand{\beq}{\begin{equation}}
\newcommand{\ber}{\begin{eqnarray}}
\newcommand{\eeq}{\end{equation}}
\newcommand{\eer}{\end{eqnarray}}

\begin{document}

\begin{titlepage}
%\vbox {\vspace{0.1mm}} %Leaves space at top of first page.
\rightline{[SUNY BING 4/22/98] }
\rightline{ hep-ph/9806373 }
\vspace{2mm}
\begin{center}
{\bf MEASUREMENT OF HELICITY PARAMETERS IN TOP QUARK DECAY}\\
\vspace{2mm}
Charles A. Nelson\footnote{Electronic address: cnelson @ 
bingvmb.cc.binghamton.edu } and Andrew M. Cohen\\
{\it Department of Physics, State University of New York at 
Binghamton\\
Binghamton, N.Y. 13902-6016}\\[2mm]
\end{center}

%\vspace{2mm}

\begin{abstract}
The standard model(SM) predicts only a $g_{V-A}$ coupling in 
$t 
\rightarrow W^+ b $ decay.  However, if additional Lorentz 
structures exist, they can manifest themselves in high energy 
processes such as this via non-SM values of the helicity 
parameters describing $t \rightarrow W^+ b $.  Plots and 
tables of the values of these helicity parameters are 
obtained for various coupling strengths.  Three phase-type
ambiguities are uncovered:  $g_{V-A} + g_{S+P} $ with 
effective-mass scale $\Lambda_{S+P} \sim -35 GeV $, $g_{V-A} 
+ g_{f_M + f_E } $ with $\Lambda_{f_M + f_E} \sim 53 GeV $, 
and an arbitrary sign-flip in the $b_L$ amplitudes $A_{X} ( 
\lambda_b = - 1/2 ) = -  A_{V-A} ( \lambda_b = - 1/2 ) $.  
These ambiguities are analyzed so that they might be 
partially resolved 
when experiments resume at the Fermilab Tevatron.
\end{abstract}

\end{titlepage}

\section{Helicity Amplitudes and $\alpha$, $\beta$, $\gamma$ 
Relative Phases}

For $t\rightarrow W^{+}b$ decay, the four on-shell helicity 
amplitudes $%
A(\lambda _{W^{+}},\lambda _b)$ can be uniquely determined by 
measurement of
four moduli and three relative phases. In Fig. 1, 
measurements in the right and left columns are respectively 
of order ${\cal O%
}(L^2)$ and  ${\cal O}(R^2)$. The interference measurements 
between the two
columns are of order  ${\cal O}(LR)$.  $L$ and $R$ denote  
the $b
$ quark's helicity $\lambda _b=\mp 1/2$. The values of 
$A(\lambda _{W^{+}},\lambda _b)$ for the standard model(SM) 
are given in the
top row of Table 1.  For the pure $V-A$ 
coupling of the
standard model, the left-handed helicity $\lambda _b=-1/2$ 
amplitudes
dominate by 1 to 2 orders of magnitude for $m_b\sim 
\;4.5GeV$.  Intrinsic and relative signs of these 
helicity
amplitudes are determined by the Jacob-Wick phase convention. 

The layout of the corners in  Fig. 1 has been chosen to 
reflect the layout in the
following probability plots for $P(W_L)$ versus $P(b_L)$ 
where 
$$
\begin{array}{c}
P(W_L)=
\mbox{Probability} \; W^{+} \; \mbox{ is longitudinally 
polarized,} \; \lambda_{W^{+}}=0 \\ 
\hspace*{-26mm} P(b_L)= \; \mbox{Probability} \; b \; \mbox{ 
is 
left-handed,} \; \lambda_b=-1/2
\end{array}
$$

The focus of this paper is on direct measurements of these 
$b_L$ amplitudes
by forthcoming experiments at the Fermilab Tevatron[1] and at 
the CERN
LHC[2] relative to the anticipated  pure $V-A$ predictions. 
If, for instance, sizable $b_R$ amplitudes were found to 
occur by these experiments, further analysis of coupling and 
phase-type ambiguities may be
warranted. Likewise, when more precise measurements become 
possible in later
experiments at the LHC or near the $t\bar t$ threshold at a 
linear $e\bar e$
collider, higher order QCD and EW corrections must be 
included, see [3].

Plots and tables of the values of the helicity parameters are 
given in terms
of a \newline ``$(V-A)$ + additional Lorentz structure''. 
Generically, 
we denote
these additional couplings by 
\begin{equation}
g_{Total} \equiv g_L+g_X \\
\end{equation}
$$
X=
\left\{ \begin{array}{ll}
X_c =  \; \mbox{chiral} = \{V+A,S\pm P,f_M\pm f_E\} \\
X_{nc} =  \; \mbox{non-chiral} = \{V,A,S,P,f_M,f_E\}. \\
\end{array}
\right. 
$$
For \hskip1em  $t \rightarrow W^+ b$, the most general 
Lorentz 
coupling[4] is $ W_\mu ^{*} J_{\bar b 
t}^\mu = 
W_\mu ^{*}\bar u_{b}\left( p\right) \Gamma ^\mu 
u_t \left(
k\right) $
where $k_t =q_W +p_b $, and
\begin{eqnarray}
\Gamma _V^\mu =g_V\gamma ^\mu +
\frac{f_M}{2\Lambda }\iota \sigma ^{\mu \nu }(k-p)_\nu   +
\frac{g_{S^{-}}}{2\Lambda }(k-p)^\mu  \nonumber \\ 
+\frac{g_S}{2\Lambda 
}(k+p)^\mu   
+%
\frac{g_{T^{+}}}{2\Lambda }\iota \sigma ^{\mu \nu }(k+p)_\nu 
\end{eqnarray}
\begin{eqnarray}
\Gamma _A^\mu =g_A\gamma ^\mu \gamma _5+
\frac{f_E}{2\Lambda }\iota \sigma ^{\mu \nu }(k-p)_\nu \gamma 
_5  
+
\frac{g_{P^{-}}}{2\Lambda }(k-p)^\mu \gamma 
_5  \nonumber \\ 
+\frac{g_P}{2\Lambda }%
(k+p)^\mu \gamma _5  +\frac{g_{T_5^{+}}}{2\Lambda }\iota 
\sigma ^{\mu \nu
}(k+p)_\nu \gamma _5
\end{eqnarray}
Some of the issues concerning the choice of a minimal subset 
of such couplings are discussed in Ref. [4].  The parameter 
$%
\Lambda_i =$ ``the effective-mass scale of new physics''.  
For $g_L = 1$ units with $g_i = 1$, the nominal size of 
$\Lambda_i$ is $\frac{m_t}{2} = 88GeV$, see below.  Lorentz 
equivalence theorems for these couplings are treated in 
Appendix B. Explicit
expressions for the $A(\lambda _{W^{+}},\lambda _b)$ in the 
case of these additional Lorentz structures are given in Ref. 
[4].

Improved theoretical treatments of effects of the $b$ quark 
mass  $m_b\sim
4.5GeV$  might also be important because of the small size of 
the $b_R$
amplitudes in the SM. In particular, as is discussed below, 
in many cases 
if an additional Lorentz structure occurs, finite $m_b$ 
effects lead to sizable ``oval shapes'' as the effective mass 
scale $\Lambda _i$ 
varies. Other
recent general analyses of effects in  $t\rightarrow W^{+}b$ 
decay
associated with new physics arising from large effective-mass 
scales $%
\Lambda _i$ are in Refs. [5-8]. 

The ``arrows'' in the upper part of Fig. 1 define the 
measurable $\alpha
,\beta ,\gamma $ relative phases between the four amplitudes. 
For instance,    
\begin{equation}
\alpha _0=\phi _0^R-\phi _0^L,\;\beta _L=\phi _{-1}^L-\phi 
_0^L,\;\gamma
_{+}=\phi _1^R-\phi _0^L\;
\end{equation}
where $A(\lambda _{W^{+}},\lambda _b)=|A|\exp (i\phi 
_{\lambda
_{W^{+}}}^{L,R})$. So for a pure $V-A$ coupling, the $\beta 
$'s 
vanish and all the $\alpha $'s and $\gamma $'s equal $+\pi $  
(or $-\pi $) to give the intrinsic minus sign of 
the standard
model's $b_R$ amplitudes.

The lower part of Fig. 1 displays the real part and imaginary 
part (primed) 
helicity
parameters corresponding to interference measurements of the 
respective
relative phases. For 
instance, c.f. Appendix C, 
\begin{equation}
\begin{array}{c}
\eta _L\equiv \frac 1\Gamma |A(-1,-\frac 12)||A(0,-\frac 
12)|\cos \beta _L
\\ 
\eta _L^{\prime }\equiv \frac 1\Gamma |A(-1,-\frac 12)||A(0,-
\frac 12)|\sin
\beta _L
\end{array}
\end{equation}
and
\begin{equation}
\eta _{L,R}=\frac 12(\eta \pm \omega )
\end{equation}
Explicit expressions for the eight $W$-polarimetry moduli and 
phase
parameters $(\Gamma ,\sigma ,\xi ,\zeta ;\eta ,\omega ,\eta 
^{\prime
},\omega ^{\prime })$ are given in Ref. [4], along with the 
inverse formulas for $\cos(\beta_{L,R}) $, $\sin(\beta_{L,R}) 
$.

By $\Lambda _b$ polarimetry[4], or some other $b$-polarimetry 
technique, it
might be possible to measure the $\alpha $ and $\gamma $ 
relative phase. In
the standard model, the two helicity parameters between the 
amplitudes with
the largest moduli are 
\begin{equation}
\begin{array}{c}
\kappa _0\equiv \frac 1\Gamma |A(0,\frac 12)||A(0,-\frac 
12)|\cos \alpha _0
\\ 
\epsilon _{+}\equiv \frac 1\Gamma |A(1,\frac 12)||A(0,-\frac 
12)|\cos \gamma
_{+}
\end{array}
\end{equation}
We will refer to $\kappa _0,\epsilon _{+}$ as the 
``$b$-polarimetry phase parameters''. From Figs. 1 other 
combinations of relative
phases/helicity-parameters are mathematically equivalent.

Unfortunately from the perspective of a complete measurement 
of the four
helicity amplitudes, the tree-level  values of $\kappa 
_0,\epsilon _{+}$ in
the SM are only about $1\%$. See the top line in both parts 
of Table 2,
which lists the $V-A$  values of the helicity parameters for 
$m_b=4.5GeV$. 
In the absence of $\tilde{T}_{FS} $ violation[4], the 
relative 
phases will be interger multiples of $\pi $ and all prime 
parameters will vanish. The prime
parameters are not directly discussed in this paper.

In Fig. 2 are two probability plots for $P(W_L)= 
\frac{1+\sigma}{2}$ 
versus $P(b_L)= \frac{1+\xi}{2}$.  The upper plot is for the 
case of a single additional chiral coupling 
$g_i$. The corners correspond to those of Fig. 1. So the dark 
rectangle of the SM, gives the relative magnitude of the 
square of the moduli of its four basic helicity amplitudes. 
Also, note from the dashed horizontal oval that an
additional $V+A$ coupling does not change the SM expectation 
that approximately $70\%$ of the final $W$'s in 
$t\rightarrow W^{+}b$
decay will be longitudinally polarized. 

The endpoints of each oval are at the dark SM rectangle and 
the dark ellipse where the coupling is pure $g_i.$ In 
general, 
the non-zero area of an oval depends 
monotonically on $m_b=4.5GeV$ and the area will increase if a 
larger value is chosen for $m_b.$ The captions to the figures 
in this paper discuss the signs of
effective-mass scales $\Lambda _i$ associated with the two 
parts of each
oval which lie between the two endpoints. Appendix A gives 
numerical values of the 
$\Lambda _i$
corresponding to various values of the helicity parameters.

The lower plot in Fig. 2 is for the case of a single 
additional 
non-chiral coupling $V,S,f_M$ ($A,P,f_E$). The corresponding 
ovals in the two non-chiral plots are almost identical in 
shape. The $g_V$ ($g_A$) endpoints lie on the upper(lower) 
parts of their ovals.

In this paper, we omit the $A,P,f_E$ curves corresponding to 
the ones provided for $V,S,f_M$ because by Lorentz invariance 
the corresponding ovals, etc., are almost identical.  The 
slight shape differences are very minor to the eye and are 
definitely negligible versus the resolutions of the 
forthcoming experiments and the omitted higher order 
theoretical contributions.  The necessary mirror reflections, 
in a few cases, to produce the omitted curves are explained 
in 
the captions to the figures.

\section{Moduli Parameters and Phase-Type Ambiguities}

Versus predictions based on the SM, two phase-type 
ambiguities arise by
consideration of the effects of a single additional 
``chiral'' coupling $g_i$
on the three moduli parameters $\sigma =P(W_L)-P(W_T),\;\xi 
=P(b_L)-P(b_R),\;
$and $\zeta =\frac 1\Gamma (\Gamma _L^{b_L-b_R}-\Gamma 
_T^{b_L-b_R})$. The
partial width $\Gamma $ for $t\rightarrow W^{+}b$ is the 
remaining and  very
important moduli parameter. However, since $\Gamma $ sets 
the overall
scale, it cannot be well measured by spin-correlation 
techniques,
which better measure the ratios of moduli and relative 
phases. So, we consider 
$\Gamma $ separately; see also [9] and references therein.

For an additional $S+P$ coupling with $\Lambda _{S+P}\sim -
34.5GeV$ the
values of $(\sigma ,\xi ,\zeta )$ and also of the partial 
width $\Gamma $
are about the same as the SM prediction, see Table 2. This is 
the first ambiguity.  The dependence of the $ \sigma $ / 
$P(W_L)$ value versus the 
effective-mass scale 
$\Lambda _{S+P}$
is shown in the upper plot in Fig. 3. Table 1 shows that this 
ambiguity will 
also occur if the
sign of the $A_X(0,-\frac 12)$ amplitude for $g_L+g_X$ is 
taken to be
opposite to that of the SM's amplitude. Recall that an 
additional $S\pm P$
only effects the longitudinal $W^{\pm}$ amplitudes and not 
the 
transverse $\lambda
_W=\mp 1$ ones. By requiring that 
\begin{equation}
\frac{A_X(0,-\frac 12)}{A_X(-1,-\frac 12)}=-\;\frac{A_L(0,-
\frac 12)}{%
A_L(-1,-\frac 12)}
\end{equation}
for $X=S+P$, we obtain%
\begin{equation}
\Lambda _{S+P}=-
(\frac{g_{S+P}}{g_L})\frac{m_t \; q_W}{2(E_W+q_W)} \sim -
(\frac{g_{S+P}}{g_L}) \frac{m_t}{4} ( 1-(\frac{m_W}{m_t})^2 
)
\end{equation}
in the notation of [4]. 

Several different definitions can be used to characterize the 
phase-type
ambiguities we consider in this section. It is important, we 
think, to
regard these ambiguities from (i) the signs in their 
$b_L$ 
amplitudes versus those for 
the SM, c.f. Table 1,  and from (ii) their 
associated additional
Lorentz structures. From this perspective, the precise 
characterization of $\Lambda
_{S+P}$, as in (8,9) for example, is not so important. 
However, if a non-SM sign were
discovered to occur in nature, further analysis and precise 
measurements would be warranted.

For an additional $f_M+f_E$ coupling with $\Lambda 
_{f_M+f_E}\sim 53GeV$ the
values of $(\sigma ,\xi ,\zeta )$  are also about the same as 
the 
SM prediction,
see Table 2. This is the second ambiguity. In this case, the 
partial width $\Gamma $ is 
about half that of
the SM due to destructive interference. The dependence of the 
\newline $\sigma $ / $%
P(W_L)$ value versus the effective-mass scale $\Lambda 
_{f_M+f_E}$ is shown
in the lower plot of Fig. 3. Table 1 shows that this 
ambiguity 
will also occur 
if the sign of
the $A_X(-1,-\frac 12)$ amplitude for $g_L+g_X$ is taken to 
be opposite to
that of the SM's amplitude. Again, from (8) for $X=f_M+f_E$, 
we obtain%
\begin{equation}
\Lambda _{f_M+f_E}= 
(\frac{g_{f_M+f_E}}{g_L})\frac{m_tE_W}{2(E_W+q_W)} \sim  
(\frac{g_{f_M+f_E}}{g_L}) \frac{m_t}{4} ( 
1+(\frac{m_W}{m_t})^2 )
\end{equation}
from Eqs.(31) in [4] since $\frac{m_b}{m_t}\frac{\sqrt{E_b-
q_W}}{\sqrt{%
E_b+q_W}}\sim 10^{-3}$. 

These phase-type ambiguities are, of course, not the same 
dynamical issue as finding a combination of $f_M + f_E$ and 
$S + P$ couplings which 
give the identical $%
b_L$ amplitudes as for a pure $V-A$ coupling. By the 
expressions in Appendix B, this is possible if $%
\Lambda _{S+P}=-\Lambda _{f_M+f_E}=\frac{m_t}2(1-
(\frac{m_b}{m_t})^2) = 87GeV$ and a negligible $%
\Lambda _{S-P}=-\Lambda _{f_M-f_E}=-\frac{(m_t)^2}{2m_b}(1-
(\frac{m_b}{m_t}%
)^2) = -3,401GeV $.  Alternatively, the fundamental $V$ 
coupling is removed by $ \Lambda_S = - \Lambda_{f_M} = (m_t + 
m_b)$ and the $A$ coupling by $ \Lambda_{f_E} = - \Lambda_P = 
(m_t - m_b)$.

Besides the $f_M+f_E$ construction of this second phase-type  
ambiguity, it should
be kept in mind that some other mechanism might produce the 
relative sign
change shown in Table 1, but without also changing the 
absolute value of the
$b_L$ amplitudes. In this case the measurement of the partial 
width  $\Gamma 
$ would not resolve the phase ambiguity. 

From consideration of Table 1, a third (phase) ambiguity can 
be 
constructed by
making an arbitrary sign-flip in the $b_L$ amplitudes, so 
$A_X(\lambda
_{W,}\lambda _b=-\frac 12)=-A_{V-A}(\lambda _{W,}\lambda _b=-
\frac 12)$,
with no corresponding sign changes in the $b_R$ amplitudes. 
Resolution of
this ambiguity will require $b$-polarimetry, c.f. [4], or 
some other amplitude
interference measurement of the overall sign of the  $b_L$ 
amplitudes relative to the $b_R$ amplitudes. 

In Figs. 4 are plotted the moduli parameters $\zeta $ 
versus $\sigma $
for the case of a single additional coupling $g_i$. The 
figures are for the case of an additional chiral 
(non-chiral) coupling.

From the perspective of possible additional Lorentz 
structures, measurement
of the partial width $\Gamma $ is an important constraint. In 
particular,
this provides a strong constraint on possible $V+A$ 
couplings, see top part
of Fig. 5, in contrast to measurement of $\sigma $ / $P(W_L)$ 
which does
not, recall Fig. 2. The remaining parts of Fig. 5 are 
for $S\pm
P$ ($f_M\pm f_E$). Likewise, as shown in the top part of 
Figs. 6,
$\Gamma $ provides a useful constraint for the possibility of 
additional $V$
and $A$ couplings which are appealing from the perspective of 
additional
gauge-theoretic structures. Here also, the lower part of 
this figure is for an additional $S,f_M$ ($P,f_E$) coupling.

\section{Phase Parameters}

In Figs. 7 are plotted the $\eta $ versus $\omega $
for the case of a single additional coupling $g_i$. The 
figures are for the case of an additional chiral 
(non-chiral) coupling. Quite dramatically in the upper plot, 
the $S+P$ and $f_M + f_E$ ambiguities both correspond to a 
``pseudo-image of the SM rectangle".  This image is in the 
third quadrant on the diagonal at $(\eta,\omega)=(-0.46,-
0.46)$.  As shown in the bottom part of Table 2, measurement 
of the signs of either of the $W$-polarimetry phase 
parameters 
$\eta $ or $\omega $ will resolve both the $S+P$ and the 
$f_M+f_E$ phase-type ambiguities.   In the SM, these
parameters are sizable and are equal if the $b_R$ amplitudes 
are omitted, see eqs.(5,6).

As discussed above, determination of the $\alpha $ and 
$\gamma $ relative
phases, as well as resolution of the third ambiguity, will
require direct empirical information about the $b_R$ 
amplitudes. One way
would be from the $b$-polarimetry phase parameters $\epsilon
_{+}$ and $\kappa_0$.  In Figs. 8 are plotted 
$\epsilon
_{+}$ versus $\eta_L$ for the
case of a single additional coupling $g_i$. The figures are  
for the case of an additional non-chiral (chiral) coupling. 
Here in general, the
non-chiral couplings produce larger values for $\kappa _0$ 
and $\epsilon _{+}
$ and so we display the non-chiral case first. In particular, 
additional $S+P$ and $f_M + f_E$ couplings have negligible 
effects on $\epsilon
_{+}$ and $\kappa_0$, see captions.

Not shown in these figures for $(\epsilon
_{+},\eta _L)$ and  
$(\kappa _0,\eta _L)$ is the unitarity limit, which is a 
circle of 
radius $\frac 12$
centered on the origin.

In Figs. 9 are plotted $\kappa_0$ versus $\eta 
_L$ for the case
of a single additional coupling $g_i$. The figures are for 
the
case of an additional non-chiral (chiral) coupling. 

\section{Ambiguities Among Other Lorentz Structures}

From the plots for the various helicity parameters, it is 
evident that there
also are ambiguities within certain subsets of the couplings 
if an
additional Lorentz structure were to occur in the form of a 
single
additional $g_i$. The 
occurrence of an
additional Lorentz structure would also raise the issue of 
how the sign of
its $\Lambda _i$ could be determined.

The following equivalence classes among additional Lorentz 
structures (versus subsets of possible experimental tests) is 
another consequence of the underlying Lorentz invariance of 
(2,3), etc.  Second, with only $W$-polarimetry, the effects 
of 
the non-zero $m_b$ mass ($m_b = 4.5GeV$) are negligible for 
(i) additional gauge couplings $V,A,V+A$ and for (ii) 
additional chiral couplings.  However, there is a sizable 
$m_b$ dependence in some chiral couplings in the $(\epsilon
_{+},\eta _L)$ and $(\kappa _0,\eta _L)$ plots.  In general 
for additional $S,P,f_M,f_E$ couplings, the dependence on 
$m_b$ is sizable and is likely to be a serious systematic 
effect, for instance in excluding possible effects from from 
fundamental or induced couplings with these Lorentz 
structures.

\subsection{Additional $V+A, V,$ or $A$ couplings}

From the gauge theory viewpoint, it is important to search 
for additional vector and axial vector couplings.  The SM's 
$P(W_L)$ and $\eta$ values are only slightly affected by 
them.  
But the values for $\xi$ (equivalently $P(b_L)$), $\zeta$, 
$\omega$, 
$\epsilon_+$, and $\kappa_0$ are significantly different from 
those of the SM.  However, inspection of the figures shows 
that in many of the plots the ovals for $V+A,V,A$ are 
approximately degenerate.  Nevertheless, from the 
different locations of their endpoints in Figs.(8-9), the 
$%
\epsilon _{+},\eta _L,\kappa _0$ parameters could be useful 
in resolving them. So $b$-polarimetry or $\Gamma $ would 
generally be
useful to resolve these additional couplings
and to
determine the sign of the associated $\Lambda _i$ .

\subsection{Additional $S - P , S,$ or $P$ couplings}

For $S,P,$ versus $S-P$ there are differences in some of the 
plots but
sufficient resolution and control of possible $m_b$ effects 
would be needed.  In particular, the narrow $S-P$ oval and 
the degenerate fat 
$S,P$ ovals lie approximately in the same $P(W_L),P(b_L)$ 
regions and also in the same $\zeta, \sigma$ regions. The 
sign 
of 
$\Lambda _i$ is the same for the $S$ and 
$P$ ovals. If  $\eta ,\omega <0$, it would
exclude  $S-P$ and would  determine the respective sign of 
$\Lambda_i$. The $\kappa _0, \eta_L$ plot is useful for 
distinguishing $S$ 
versus $P$ and for the sign of $\Lambda _i$. If $S-P$ were 
resolved, then $\kappa _0$ would give the sign of 
$\Lambda _i$. $\Gamma $ is not useful for 
separating $S$ versus $P$, but $%
\Gamma $ is different for $S-P$.

\subsection{Additional $f_M+f_E$ or $S+P$ couplings}

$f_M+f_E$ and $S+P$ can be distinguished from either the 
$P(W_L),P(b_L)$ or $%
\zeta ,\sigma $ plots. Once separated, $\Gamma $ could 
provide information
on the sign of $\Lambda _i$. If $\eta ,\omega <0$, it would 
determine the respective sign of $\Lambda _i$.  $\epsilon_{+}  
\simeq \kappa_0 \simeq 0$ for these couplings. 

\subsection{Additional $f_M-f_E$ , $f_M$, or $f_E$ couplings}

With sufficient resolution and control of $m_b$ effects, 
$f_M-f_E$ could be separated versus 
$f_M$, $f_E$ by $%
P(W_L),P(b_L)$; by the $\zeta ,\sigma $ plot; and/or by 
$\Gamma $.
The $\epsilon _{+}$,$\eta _L$ plot would be useful for 
separating $f_M$ from 
$f_E$ and in determining the sign of $\Lambda _i.$ It would 
also determine
the sign for $f_M-f_E$. 

\begin{center}
{\bf Acknowledgments}
\end{center}

One of us(CAN) thanks the Fermilab Theory Group for a useful 
and intellectually stimulating visit.  For computer services,  
we thank Mark Stephens.  This work was partially supported by 
U.S. Dept. of Energy Contract No. DE-FG 02-96ER40291.

\newpage

\begin{center}
{\bf Appendices}
\end{center}

\appendix

\section{Effective-mass scales $\Lambda_i$ corresponding to 
\newline values of the helicity parameters}

The preceding plots do not quantitatively display the 
effective-mass scales 
$\Lambda _i$ associated with the ovals for the respective 
$g_{Total}=g_L+g_X$
Lorentz structures. Instead, we present this information in 
Tables 3-6. Note that $P(W_L)=\frac{1+\sigma }2$ and $%
P(b_L)=\frac{1+\xi }2$. 

\section{Lorentz equivalence theorems}

In the case of non-chiral couplings and with the signs and 
normalizations of
(2,3), the tensorial  $f_M$ coupling can be absorbed by using 
\begin{equation}
g_V^{\prime }=g_V-(m_t+m_b)\frac{f_M}{2\Lambda 
_M},\;\;\frac{g_S^{\prime }}{%
2\Lambda _S^{\prime }}=\frac{g_S}{2\Lambda 
_S}+\frac{f_M}{2\Lambda _M},
\end{equation}
or alternatively, the scalar $S$ coupling can be absorbed
\begin{equation}
g_V^{\prime }=g_V+(m_t+m_b)\frac{g_S}{2\Lambda 
_S},\;\;\frac{f_M^{\prime }}{%
2\Lambda _M^{\prime }}=\frac{f_M}{2\Lambda 
_M}+\frac{g_S}{2\Lambda _S}.
\end{equation}
Similarly, $f_E$ can be absorbed by 
\begin{equation}
g_A^{\prime }=g_A+(m_t-m_b)\frac{f_E}{2\Lambda 
_E},\;\;\frac{g_P^{\prime }}{%
2\Lambda _P^{\prime }}=\frac{g_P}{2\Lambda 
_P}+\frac{f_E}{2\Lambda _E},
\end{equation}
or alternatively $P$ by 
\begin{equation}
g_A^{\prime }=g_A-(m_t-m_b)\frac{g_P}{2\Lambda 
_P},\;\;\frac{f_E^{\prime }}{%
2\Lambda _E^{\prime }}=\frac{f_E}{2\Lambda 
_E}+\frac{g_P}{2\Lambda _P}.
\end{equation}
The $g_{T^{+}}$ is absorbed by $g_V\rightarrow g_V^{\prime 
}=g_V-(m_t-m_b)%
\frac{g_{T^{+}}}{2\Lambda _{T^{+}}}$ and $g_{T_5^{+}}$ by 
$g_A\rightarrow
g_A^{\prime }=g_A+(m_t+m_b)\frac{g_{T_5^{+}}}{2\Lambda 
_{T_5^{+}}}.$

In the case of the chiral combinations, the tensorial $g_{\pm 
}\equiv $ $%
f_M\pm f_E$ are absorbed by using 
\begin{equation}
\begin{array}{c}
g_L^{\prime }=g_L-m_t
\frac{g_{+}}{2\Lambda _{+}}-m_b\frac{g_{-}}{2\Lambda _{-
}},\;\;\frac{%
g_{S+P}^{\prime }}{2\Lambda _{S+P}^{\prime 
}}=\frac{g_{S+P}}{2\Lambda _{S+P}}%
+\frac{g_{+}}{2\Lambda _{+}}, \\ g_R^{\prime }=g_R-
m_t\frac{g_{-}}{2\Lambda
_{-}}-m_b\frac{g_{+}}{2\Lambda _{+}},\;\;\frac{g_{S-
P}^{\prime }}{2\Lambda
_{S-P}^{\prime }}=\frac{g_{S-P}}{2\Lambda _{S-P}}+\frac{g_{-
}}{2\Lambda _{-}}%
,
\end{array}
\end{equation}
or alternatively $S\pm P$ by
\begin{equation}
\begin{array}{c}
g_L^{\prime }=g_L+m_t
\frac{g_{S+P}}{2\Lambda _{S+P}}+m_b\frac{g_{S-P}}{2\Lambda 
_{S-P}},\;\;\frac{%
g_{+}^{\prime }}{2\Lambda _{+}^{\prime 
}}=\frac{g_{+}}{2\Lambda _{+}}+\frac{%
g_{S+P}}{2\Lambda _{S+P}}, \\ g_R^{\prime 
}=g_R+m_t\frac{g_{S-P}}{2\Lambda
_{S-P}}+m_b\frac{g_{S+P}}{2\Lambda _{S+P}},\;\;\frac{g_{-
}^{\prime }}{%
2\Lambda _{-}^{\prime }}=\frac{g_{-}}{2\Lambda _{-
}}+\frac{g_{S-P}}{2\Lambda
_{S-P}}.
\end{array}
\end{equation}
The $\tilde g_{\pm } = g_{T^+} \pm g_{T_5^{+}}$ are absorbed 
by $g_L\rightarrow 
g_L^{\prime }=g_L-m_t%
\frac{\tilde g_{+}}{2\tilde \Lambda _{+}}+m_b\frac{\tilde 
g_{-}}{2\tilde
\Lambda _{-}}$ and $g_R\rightarrow g_R^{\prime }=g_R-
m_t\frac{\tilde g_{-}}{%
2\tilde \Lambda _{-}}+m_b\frac{\tilde g_{+}}{2\tilde \Lambda 
_{+}}$

\section{Formulas for $\alpha, \beta, \gamma$ phases from 
helicity parameters}
 
Eqs.(5,6) define the $\eta _{L,R}$ helicity parameters 
associated with the $%
\beta _{L,R}$ phases. Similarly, from Figs. 1 the 
parameters associated 
with the $\alpha _{0,1}$ and $\gamma _{\pm }$ phases are
\begin{equation}
\begin{array}{c}
\kappa _0=\frac 12(\lambda +\kappa )\equiv \frac 1\Gamma 
|A(0,-\frac
12)||A(0,\frac 12)|\cos \alpha _0 \\ 
\kappa _1=\frac 12(\lambda -\kappa )\equiv \frac 1\Gamma |A(-
1,-\frac
12)||A(1,\frac 12)|\cos \alpha _1 \\ 
\epsilon _{+}=\frac 12(\delta +\epsilon )\equiv \frac 1\Gamma 
|A(1,\frac
12)||A(0,-\frac 12)|\cos \gamma _{+} \\ 
\epsilon _{-}=\frac 12(\delta -\epsilon )\equiv \frac 1\Gamma 
|A(-1,-\frac
12)||A(0,\frac 12)|\cos \gamma _{-}
\end{array}
\end{equation}
The corresponding primed parameters are defined by replacing 
the cosine by
sine.

The inverse formulas for $\cos \beta _{L,R}$, $\sin \beta 
_{L,R}$ from $\eta
_{L,R}$ and $\eta _{L,R}^{\prime }$ are given by Eqs. (56-59) 
in [4]. For
extracting the $\alpha _{0,1}$ and $\gamma _{\pm }$ phases, 
\begin{equation}
\begin{array}{c}
\cos \alpha _0=
\frac{4\kappa _0}{\sqrt{(1+\sigma )^2-(\xi +\zeta )^2}} \\ 
\cos \alpha _1=
\frac{4\kappa _1}{\sqrt{(1-\sigma )^2-(\xi -\zeta )^2}} \\ 
\cos \gamma _{+}=
\frac{4\epsilon _{+}}{\sqrt{(1+\zeta )^2-(\sigma +\xi )^2}} 
\\ \cos \gamma
_{-}=\frac{4\epsilon _{-}}{\sqrt{(1-\zeta )^2-(\sigma -\xi 
)^2}}
\end{array}
\end{equation}
and the sine's of the respective angles are obtained by using 
the primed
helicity parameter in the respective numerator.

\newpage

\begin{center}
{\bf Table Captions}
\end{center}

Table 1: For the $ ( \sigma, \xi, \zeta ) $ ambiguous-moduli 
points, numerical values of the associated helicity 
amplitudes $ A\left( \lambda_{W^{+} } ,\lambda_b \right) $.  
The values for the amplitudes are listed first in $ g_L = 1 $ 
units, and second as $ A_{new} = A_{g_L = 1} / \surd \Gamma $ 
which removes the effect of the differing partial width, $ 
\Gamma $ for $ t\rightarrow W^{+}b $. [$m_t=175GeV, \; m_W = 
80.35GeV, \; m_b = 4.5GeV$ ]. 

Table 2:  For the $ ( \sigma, \xi, \zeta ) $ ambiguous-moduli  
points, numerical values of the associated helicity 
parameters.  Listed first are the four moduli parameters.  
Listed second are the values of phase parameters which could  
be used to resolve the ambiguities.

Table 3: Numerical values of the associated helicity 
parameters $ \sigma, \xi, 
\zeta, \eta, $ and $ \omega $ as the effective-mass scales $ 
\Lambda_i $ for additional chiral Lorentz structures vary 
over the range $ ( -5000 GeV, 5000 GeV ) $.  
For an additional $ V+A $ coupling, $g_R$ varies over the 
range $ (-1.6667, 1.6667) $ with fixed $g_L =1$. 
 
Table 4: Numerical values 
of the associated helicity parameters $ \sigma, \xi, \zeta, 
\eta, $ and $ \omega $ as the effective-mass scales $ 
\Lambda_i $ for additional non-chiral Lorentz structures vary 
over the range $ ( -5000 GeV, 5000 GeV ) $.  
For an additional $ V $ or $ A $ coupling, $g_{V,A}$ 
respectively varies over the range $ (-1.6667, 1.6667) $ with 
fixed $g_L =1$. 

Table 5: Numerical 
values of the associated helicity parameters $ \Gamma, 
\eta_L, \kappa_0, $ and $ \epsilon_{+} $ as the effective-
mass 
scales $ \Lambda_i $ for additional chiral Lorentz structures 
vary over the range $ ( -5000 GeV, 5000 
GeV ) $.  For an additional $ V+A $ coupling, $g_R$ varies 
over the range $ (-1.6667, 1.6667) $ with fixed $g_L =1$. 
 
Table 6: Numerical values 
of the associated helicity parameters $ \Gamma, \eta_L, 
\kappa_0, $ and $ \epsilon_{+} $ as the effective-mass scales 
$ 
\Lambda_i $ for non-chiral Lorentz structures vary over the 
range $ ( -5000 GeV, 5000 GeV ) $.  
For an additional $ V $ or $ A $ coupling, $g_{V,A}$ 
respectively varies over the range $ (-1.6667, 1.6667) $ with 
fixed $g_L =1$. 

\newpage

\begin{center}
{\bf Figure Captions}
\end{center}

FIG. 1: For $ t\rightarrow W^{+}b $ decay, display of the 
four helicity amplitudes $ A\left( \lambda 
_{W^{+} } ,\lambda _b \right) $ relative to the b quark's 
helicity.  The upper sketch defines the measurable `` $ 
\alpha, \beta, \gamma $ "  
relative phases, c.f. Eqs(4). The lower sketch defines the 
real part and imaginary part
(primed) helicity parameters corresponding to these relative 
phases.

FIG. 2: For the case of a single additional coupling 
($g_i$), plots of the probability, $P(W_L)$, that the emitted 
$ W^{+} $ 
is ``Longitudinally" polarized versus the probability, 
$P(b_L)$, that the 
emitted b-quark has ``Left-handed" helicity.  The upper plot 
is for additional chiral couplings:  a dark 
rectangle denotes the value for the pure $ V - A $ coupling 
of 
the standard model. The long-dashed (horizontal) oval is for 
an additional $ V+A $ coupling.  A dark ellipse denotes the 
end point where the coupling is pure $ V + A $, and similarly 
for the other ovals.  The dashed oval is for an additional $ 
f_M - f_E $ coupling.  
The dashed-dot oval is for an addition $ S - P $ coupling. 
The 
solid (zero-area) vertical ovals 
with $ 
P(b_L) = 1 $ which end above/below the $ V-A$ 
point are for an additional $ f_M + f_E $ / $ 
S+P $ coupling. 
The upper(lower) portions of the ovals are for  
$ \Lambda_i > 0 ( < 0 )$, except for the solid curves $ f_M + 
f_E $ 
and $ S+P $ which cover the full $P(W_L)$ range for small  $ 
\Lambda_i $ values, see the $P(W_L)$ versus $\Lambda_i $ 
plots
in Figs. 3. The lower plot is for additional non-chiral 
coupling $V, S, f_M $ couplings.  The long-dashed 
(horizontal) 
oval is for an 
additional $ V(A)$  coupling.  The dashed oval is for an 
additional $ f_M(f_E) $ coupling.  The dashed-dot oval is for 
an addition $ 
S (P) $ coupling.  $ \Lambda_i > 0 $ corresponds to the tops 
of the ovals from the $V - A$ solid rectangle to the pure 
$g_i$ endpoints.  To the eye, the omitted (see Sec. 1)  
respective curves 
for $A,P,f_E$ almost overlap the ones for $V,S,f_M$.  For 
$A$, 
the endpoint is 
slightly below (that for $V$) and on the bottom arc of its 
oval.

FIG. 3: The upper(lower) plot displays the $P 
(W_L) $ value versus the effective-mass scale $ 
\Lambda $ for an 
additional $ S+P$ ($f_M + f_E$) coupling.  The ambiguous-
moduli point for 
this coupling occurs at $ \Lambda_{S+P} \sim -34.5 GeV $ ($ 
\Lambda_{f_M + f_E } \sim 
52.9 GeV $ )
where 
the solid curve crosses over the dashed horizontal line which 
shows the standard V-A value.  

FIG. 4: For the case of a single additional coupling 
($g_i$), plots of the moduli parameters $\zeta$ versus 
$\sigma$. The ovals are labeled as in Fig. 2.  $ \Lambda_i > 
0 $ corresponds to the right-sides of the ovals from the $V - 
A$ rectangle to the pure $g_i$ endpoints.  In the upper plot 
for additional chiral couplings, the $S+P$ ($f_M + f_E$) 
endpoint is in the first(third) quadrant.  The lower plot is 
for additional $V, S, f_M $ couplings.  To the eye, the 
omitted respective curves for $A, P, f_E $ almost overlap.  
The $A$ endpoint is slightly to the left, on the origin side 
of the oval. 

FIG. 5: Plots of the partial width for $ t\rightarrow W^{+}b 
$ versus strengths of an additional chiral coupling: 
upper-figure is for an additional $ V+A$ coupling; middle-
figure's solid ( dashed-dot) curve is for 
$S+P$ ($S-P$ ); and lower-figure's solid ( dashed-dot) curve 
is for $f_M + f_E$ ($ f_M - f_E$ ).  

FIG. 6: Plots of the partial width for $ t\rightarrow W^{+}b 
$ versus strengths of an additional non-chiral coupling $V, 
S, f_M$: 
upper-figure is for an additional $ V$ coupling; and lower-
figure's dashed-dot ( dotted ) curve is 
for $S$ ($ f_M $ ).  The omitted plot for $A$ is almost the 
mirror image about the $\Gamma$ axis of $V$'s, so $\Gamma 
(g_A) \approx \Gamma ( - g_V ) $.  Those for $P,f_E$ are 
respectively about the same as for $S,f_M$.

FIG. 7: Plots of the two W-polarimetry phase parameters, 
$\eta, \omega $ for the case of a single additional coupling 
($g_i$).   The ovals are labeled as in Fig. 2.  $ 
\Lambda_i > 0  $ correspond to the lower parts of the ovals 
from the $V - A$ rectangle to the pure $g_i$ endpoints. The 
upper plot is for additional chiral couplings and the first 
two phase-type ambiguities correspond to a ``pseudo-image of 
the SM rectangle" at $(-0.46,-0.46)$.  The $S \pm P $ end 
points are at the origin.  On the solid 
line (zero area) ovals, the $ f_M + f_E $ end point is in the 
first quadrant, and both the $ f_M + f_E $ and $S+P$ ovals
extend through the third quadrant with respectively 
$\Lambda_i  
> 0, < 0 $. The $f_M + f_E$ and $S+P$ ovals each cover the 
entire diagonal. The lower plot is for additional $V, S, f_M 
$ 
( $A, P, f_E $) couplings. On the vertical axis, the $V(A)$ 
end 
point is at the bottom(top) of its horizontal 
oval. Similarly, near the origin the $S(P)$ end point is 
at the bottom(top) of its oval.  

FIG. 8: Plots of the b-polarimetry phase parameter 
$\epsilon_+$ versus $\eta_L $ for the case of a single 
additional coupling ($g_i$).  The upper plot is for 
additional 
$V, S, f_M $ couplings.  $ \Lambda_i > 0  $ 
corresponds to the upper part of the long-dashed $V$ oval 
from the $V - A$ rectangle, the lower part of the dotted 
$f_M$, and the positive $\eta_L$ part of the dashed-dot $S$.  
The latter, zero-area $S$ oval extends to $ \eta_L \sim -
1.2$.  
The omitted $A, P, f_E $ plot is almost the mirror image 
about 
the $\eta_L$ axis, in it $ \Lambda_i > 0  $ corresponds to 
the 
upper parts of the long-dashed $A$ and dotted $f_E$ ovals 
from the $V - A$ rectangle, and the positive $\eta_L$ part of 
the dashed-dot $P$.  The lower plot is for a single 
additional chiral coupling.  Only the $f_M + f_E$ endpoint is 
not near the origin. The $\epsilon_+$ values are non-
negligible for only two couplings:  $ \Lambda_i > 0  $ 
corresponds to the upper part of the long-dashed $V+A$ oval 
from the $V - A$ rectangle, the lower part of the dotted 
$f_M-
f_E$ oval.  For the other couplings, their $\eta_L = \eta + 
\omega$ dependence is shown in Fig. 7.

FIG. 9: Plots of the b-polarimetry phase parameter 
$\kappa_0$ versus $\eta_L $ for the case of a single 
additional ($g_i$).  
The upper plot is for additional $V, 
S, f_M $ couplings.  $ \Lambda_i > 0  $ corresponds to the 
upper part of the long-dashed $V$ and the dashed-dot $S$ 
ovals 
from the $V - A$ rectangle, and corresponds to the lower part 
of the dotted $f_M$.  The omitted $A, P, f_E $ plot is almost 
the mirror image about the $\eta_L$ axis.  In it, $ \Lambda_i 
> 
0  $ corresponds to the upper parts of the long-dashed $A$ 
and 
dotted $f_E$ ovals from the $V - A$ rectangle, and 
corresponds 
to the lower part part of the dashed-dot $P$. The lower plot  
is for a single additional chiral coupling. Only the $f_M + 
f_E$ endpoint is not near the origin.  The $\kappa _0$ values 
are non-negligible for three couplings:  $ \Lambda_i > 0  $ 
corresponds to the upper part of the long-dashed $V+A$ and 
dashed-dot $S-P$ ovals from the $V - A$ rectangle, and 
corresponds to the lower part of the dotted $f_M-f_E$ oval.  

\end{document}